\documentclass[pra,amsmath,amssymb,showpacs,superscriptaddress, twocolumn]{revtex4}
\usepackage[english]{babel}
\usepackage[latin1]{inputenc}
\usepackage{hyperref}
\usepackage{amsmath, amssymb, amsfonts}
\usepackage{amssymb,xcolor}
\usepackage{graphicx}
\graphicspath{{images/}}
\usepackage{exscale}
\usepackage{graphicx}
\usepackage{amsmath}
\usepackage{latexsym}
\usepackage{amsfonts}
\usepackage{amssymb}

\newcommand{\ket}[1]{|#1\rangle}
\newcommand{\bra}[1]{\langle #1|}
\newcommand{\Tr}{\mathrm{Tr}}
\newcommand{\ud}{\mathrm{d}}
\newcommand{\Cr}{\mathcal{C}_{r}}
\newcommand{\Cl}{\mathcal{C}_{l_1}}
\newcommand{\D}{\mathcal{D}}
\newcommand{\B}{\mathcal{B}}
\newcommand{\I}{\mathcal{I}}
\newcommand{\C}{\mathcal{C}}
\newcommand{\n}{\nonumber\\}

\begin{document}
\title{Coherence depletion in the Grover quantum search algorithm}
\author{Hai-Long Shi}
\affiliation{Institute of Modern Physics, Northwest University, Xi'an 710069, China}
\affiliation{School of Physics, Northwest University, Xi'an 710069, China}
\author{Si-Yuan Liu}\email{syliu@iphy.ac.cn}
\affiliation{Institute of Modern Physics, Northwest University, Xi'an
710069, China}
\affiliation{Shaanxi Key Laboratory for Theoretical Physics Frontiers, Xi'an 710069, China}
\author{Xiao-Hui Wang}
\affiliation{School of Physics, Northwest University, Xi'an 710069, China}
\affiliation{Shaanxi Key Laboratory for Theoretical Physics Frontiers, Xi'an 710069, China}
\author{Wen-Li Yang}
\affiliation{Institute of Modern Physics, Northwest University, Xi'an
710069, China}
\affiliation{Shaanxi Key Laboratory for Theoretical Physics Frontiers, Xi'an 710069, China}
\author{Zhan-Ying Yang}
\affiliation{School of Physics, Northwest University, Xi'an 710069, China}
\affiliation{Shaanxi Key Laboratory for Theoretical Physics Frontiers, Xi'an 710069, China}
\author{Heng Fan}
\affiliation{Institute of Physics, Chinese Academy of Sciences, Beijing
100190, China}
\affiliation{Institute of Modern Physics, Northwest University, Xi'an
710069, China}
\affiliation{Shaanxi Key Laboratory for Theoretical Physics Frontiers, Xi'an 710069, China}

\begin{abstract}
We investigate the role of quantum coherence depletion (QCD) in Grover search algorithm (GA) by using several typical measures of quantum coherence and
quantum correlations.
By using the relative entropy of coherence measure ($\Cr$), we show that the success probability depends on the QCD.
The same phenomenon is also found by using the $l_1$ norm of coherence measure ($\Cl$).
In the limit case, the cost performance is defined to characterize the behavior about QCD in enhancing the success probability of GA,
which is only related to the number of searcher items and the scale of database, no matter using $\Cr$ or $\Cl$.
 In generalized Grover search algorithm (GGA), the QCD for a class of states increases with the required optimal measurement time.
 In comparison, the quantification of other quantum correlations in GA,
 such as pairwise entanglement, multipartite entanglement, pairwise discord and genuine multipartite discord,
cannot be directly related to the success probability or the optimal measurement time.
 Additionally, we do not detect pairwise nonlocality or genuine tripartite nonlocality in GA since
Clauser-Horne-Shimony-Holt inequality and Svetlichny's inequality are not violated.

\end{abstract}





\maketitle
\section{Introduction}

Quantum mechanics provides some distinctive computational resources that can be utilized to make quantum algorithms
superior to some classical algorithms \cite{Nielsen}.
The origin of this speed-up in quantum computational processes has attracted many research attentions.
For instance, Jozsa and Linden demonstrated that, for pure states,
entanglement is need for some certain quantum computations if the calculated results can not be simulated classically \cite{Role-Ent1}.
In addition, Vidal showed that, under arbitrary bipartite cut and at all times, if the state of the quantum computer has Schmidt rank polynomial in $n$
then the quantum computation can be simulated classically \cite{Role-Ent2}.
However, a quantum computation using only separable states still surpasses classical computations \cite{Role-Ent2}.
The celebrated Knill-Gottesman theorem tells us
that some quantum algorithms using highly entangled states can also be efficiently simulated classically \cite{Role-Ent3}.
Thus, the existence of entanglement is not sufficient for exponential quantum speed-up \cite{RMP-Ent}.
Besides entanglement, quantum discord, as another type of quantum correlations, is equally vital in quantum algorithms.
For example, in the some settings of one-way algorithm for remote state preparation,
discord does not vanish while entanglement vanishes
,when the noise is maximal and fidelity drops to its minimum value
\cite{Role-Dis}.
Moreover, the effects of quantum resources, such as entanglement, discord and nonlocality on the process of quantum key distribution (QKD)
receive widespread attention and scrutiny \cite{QKD1, QKD2, QKD3}.

Coherence, as a quantum property from the quantum states superposition principle \cite{Schrodinger},
has been widely studied in quantum information processing \cite{Bagan, Jha, Kammerlander}.
A rigorous framework for quantifying the coherence was proposed by Baumgratz \emph{et al.} in Ref. \cite{Baumgratz}.
Recently, coherence has been proved that it can be converted to other valued quantum resources,
such as entanglement and discord, by suitable operations \cite{covert1, covert2, covert3}.
To some extent, coherence is the same important as well as entanglement or discord.
Moreover, coherence also exists in a single system without any correlations.
A natural question is what the role of coherence plays in quantum algorithms?

Recently, this topic has generated a great deal of interest.
Hillery declared that coherence can be viewed as a resource in Deutsch-Jozsa algorithm in the sense
that a bigger amount of coherence decreases the failure of this algorithm \cite{Role-Coher1}.
For deterministic quantum computation with one qubit (DQC1),
Matera \emph{et al.} displayed that the precision of this algorithm is directly related to the recoverable coherence \cite{DQC1, Role-Coher2}.
At the heart of quantum algorithms, there lies another fundamental algorithm, Grover search algorithm (GA) \cite{Grover, Galindo}.
GA was introduced for accelerating search process \cite{Childs}.
It is believed that multipartite entanglement is necessary for GA to achieve the speed-up \cite{Role-Ent1}.
To investigated properties of entanglement, different measures of entanglement, such as concurrence and geometric measure of entanglement,
have been attempted in GA \cite{Fang, Meyer, Macchiavello, Rossi, Rungta, Chakraborty}.
However, the role of entanglement is not yet fully demonstrated, in particular,
the quantity of entanglement is not directly related with the success probability in GA \cite{Braunstein}.
On the other hand, quantum discord, as a nonclassical correlation beyond entanglement, has been proved that its behavior is similar to the entanglement in GA \cite{Cui}.
It is worth noting that coherence is potentially a more fundamental quantum resource
than entanglement and discord \cite{Sun}. Much attention has been paid
in this direction \cite{plenioreview,yangdong,hengfan1,hengfan2,hengfan3,hengfan4}.
Whether or not coherence will display unique characteristics, which are different from entanglement or discord in GA?
To clarify the role of coherence, we investigate quantum coherence depletion (QCD) in GA and in generalized Grover search algorithm (GGA).
Other quantum correlations are also discussed for comparison.

This paper is organized as follows.
In Sec. \ref{section1}, we briefly review GA
and study its coherence dynamics of the whole $n$-qubit system in the cases of any solutions to the search problem
by using two different coherence measures, namely, the relative entropy and the $l_1$ norm.
In additional, the relationship between QCD and success probability in GA is also discussed.
In Sec. \ref{section2}, we introduce GGA and investigate the relationship between QCD of a class of states and optimal measurement time in GGA.
In Sec. \ref{section3}, we consider dynamics of entanglement, discord and nonlocality for any two qubits in the simplest situation of  single solution to Grover search. Moreover, multipartite entanglement, genuine quantum correlation and genuine tripartite nonlocality are also discussed.
Finally, the main results are summarized in Sec. \ref{section5}.

\section{COHERENCE DEPLETION IN STANDARD GROVER SEARCH ALGORITHM}\label{section1}
The first step of GA is to initialize the $n$-qubit database to an equally weighted superposition of all computational basis states $\ket{\psi_0}=1/\sqrt{2^n}\sum^{2^n-1}_{x=0}\ket{x}$,
which can be realized by projecting a prepared pure state $\ket{0,...,0}$ to local Hadamard gates $H^{\otimes n}$
where $H=(\ket{0}\bra{0}+\ket{0}\bra{1}+\ket{1}\bra{0}-\ket{1}\bra{1})/\sqrt{2}$.
It should be pointed that the initialized $n$-qubit database is a maximally coherent state with $N=2^n$ equiprobable items $\ket{x}$
and our goal is to obtain desired items
(in the following we call them as the ``solutions" to the GA) from it with maximum probability after the GA.
The initialized database can be written in a more convenient form
\begin{equation}
\ket{\psi_0}=\sqrt{\frac{j}{N}}\ket{X}+\sqrt{\frac{N-j}{N}}\ket{X^{\perp}},
\end{equation}
where $j$ represents the number of solutions
and $\ket{X}=1/\sqrt{j}\sum_{x_s}\ket{x_s}$ [$\ket{X^{\perp}}=1/\sqrt{N-j}\sum_{x_n}\ket{x_n}$] is constructed by states $\ket{x_s}$ [$\ket{x_n}$] that are solutions [non-solutions] to the GA.
It is easy to confirm that both $\ket{X}$ and $\ket{X^{\perp}}$ are orthonormal.
The next step is to apply Grover operation $G$ repeatedly (called iteration) to improve proportion of solutions gradually.
The Grover operation, $G=DO$, is comprised of oracle $O=I-2\ket{X}\bra{X}$
and an inversion about average operation $D=2\ket{\psi_0}\bra{\psi_0}-I$ \cite{Grover}.
After $r$ iterations of the Grover operation $G$, the global state has the following form \cite{Cui, Nielsen}
\begin{equation}\label{psi-r}
\ket{\psi_r}\equiv G^r\ket{\psi_0}=\sin\alpha_r\ket{X}+\cos\alpha_r\ket{X^{\perp}},
\end{equation}
with $\alpha_r=(r+1/2)\alpha$ and $\alpha=2\arctan\sqrt{j/(N-j)}$.
Note that $\ket{\psi_r}$ is also a pure state since $G$ is unitary and initial state $\ket{\psi_0}$ is a pure state.
The above processes are summarized in Fig. 1.
\\(1) Initialize the $n$-qubit database to $\ket{\psi_0}$;
\\(2) Oracle $O$ reflects the vector $\ket{\psi_0}$ according to $\ket{X^\perp}$ and then operation $D$ reflects the vector $O\ket{\psi_0}$ according to $\ket{\psi_0}$. Therefore, the role of Grover operation $G$ is to rotate the vector before iteration anticlockwise by an angle $\alpha$.
\begin{figure}[ht]
 \centering
 \includegraphics[height=6cm]{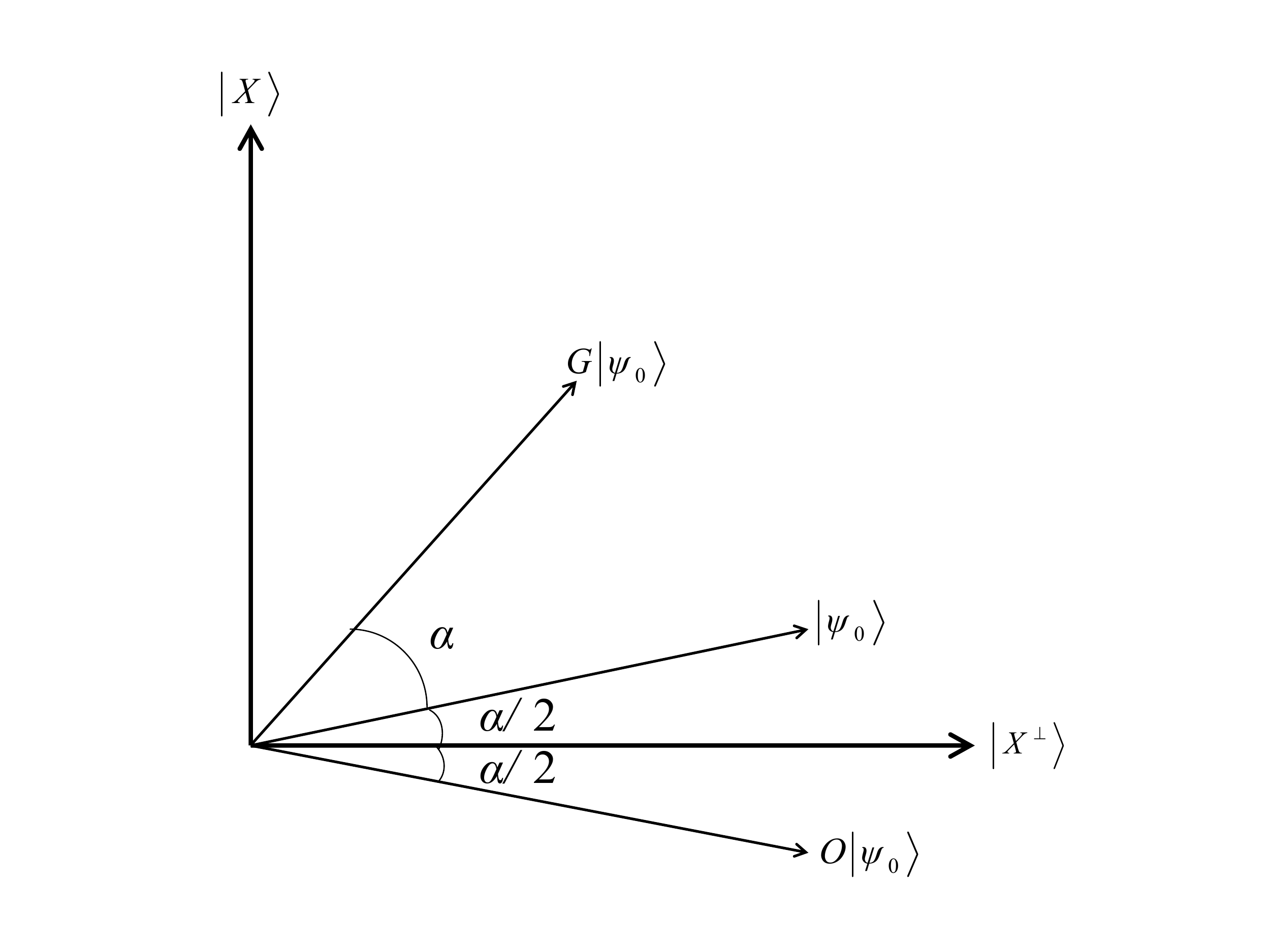}
 \caption{An illustration to show that the first two steps of GA. Firstly, initialize the $n$-qubit database to $\ket{\psi_0}$; Secondly, $O$ reflects $\ket{\psi_0}$ according to $\ket{X^\perp}$ and $D$ reflects $O\ket{\psi_0}$ according to $\ket{\psi}$. Consequently, one whole iteration $G$ turns  the vector before iteration anticlockwise by an angle $\alpha$. }
 \label{fig:algorithm}
\end{figure}

The final step (3) is that measure $\ket{\psi_r}$ to get $\ket{X}$ with maximum probability.
The success probability is expressed as
\begin{equation}\label{P}
P(r)=\sin^2\alpha_r.
\end{equation}
Therefore, the optimal time to stop iteration is $r_{opt}=CI[(\pi-\alpha)/(2\alpha)]$ where $CI[x]$ denotes the closest integer to $x$.
In the following, we will confine our discussion to $0\leq r\leq r_{opt}$.

Quantum coherence describes the capability of a quantum state to exhibit quantum interference phenomena.
The first rigorous framework to quantify the coherence was built by Baumgratz \emph{et al.} in Ref. \cite{Baumgratz}.
Based on this work, a number of coherence measures,
such as the relative entropy of coherence, the $l_1$ norm of coherence, the Tsallis relative $\alpha$ entropy of coherence and the coherence of formation \cite{Baumgratz, Yuan, Rastegin},
have been proposed.
Recently, a novel phenomenon has been founded  in Ref. \cite{Yu} that all measures of coherence are frozen for an initial state in a strictly incoherent channel if and only if the relative entropy of coherence is frozen for the state.
It means that the relative entropy of coherence is an excellent coherence measure.
Hence we choose it to investigate the GA and also calculate the $l_1$ norm of coherence for comparison.
In this section, we consider coherence dynamics under the general case of any $j$ solutions.
According to Eq. (\ref{psi-r}), the density matrix of state generated by GA can be written as
\begin{eqnarray}\label{rho-r}
\rho(r)=\frac{a^2}{j}\sum_{x_s,y_s}\ket{x_s}\bra{y_s}+b^2\sum_{x_n,y_n}\ket{x_n}\bra{y_n}\n
\quad+\frac{ab}{\sqrt{j}}[\sum_{x_s}\sum_{y_n}(\ket{x_s}\bra{y_n}+\ket{y_n}\bra{x_s})],
\end{eqnarray}
where subscripts $s$ and $n$ denote that they are solutions and non-solutions, respectively.
Here $a=\sin\alpha_r$ and $b=1/\sqrt{N-j}\cos\alpha_r$ are brought in for convenience.

\subsection{The relative entropy of coherence}
The definition of relative entropy of coherence is \cite{Baumgratz}
\begin{eqnarray}
\Cr({\rho})=\min_{\delta\in \mathcal {I}}S(\rho\|\delta),
\end{eqnarray}
where $S(\rho\|\delta)=\Tr(\rho\log_2\rho-\rho\log_2\delta)$ is the quantum relative entropy
and $\mathcal {I}$ denotes a set of incoherent quantum states whose density matrices are diagonal in the calculational basis.
This formula can be rewritten as a closed form \cite{Baumgratz}, avoiding the minimization
\begin{equation}\label{C-r}
\Cr({\rho})=S(\rho_{diag})-S(\rho),
\end{equation}
where $\rho_{diag}=\sum_i\rho_{ii}\ket{i}\bra{i}$ and $S(\rho)=-\Tr(\rho\log_2\rho)$ is the von Neumann entropy.

Substitute the Eq. (\ref{rho-r}) into Eq. (\ref{C-r}), we obtain the coherence dynamics of $n$-qubit
\begin{equation}\label{C-r-g}
\Cr(\rho)=H(a^2)+\log_2(N-j)+a^2\log_2\frac{j}{N-j},
\end{equation}
where $H(x)=-x\log_2x-(1-x)\log_2(1-x)$ is the binary Shannon entropy function.
Note that the relative entropy of coherence is independent of the choices of solutions.
In other words, it only depends on the number of solutions $j$
since $S(\rho)=0$  and $S(\rho_{diag})$ is only connected with the diagonal elements of $\rho(r)$.
From Eq. (\ref{C-r-g}), we have
\begin{equation}\label{dC-re/dr}
\frac{\ud\Cr(\rho)}{\ud r}=\log_2\frac{j(1-a^2)}{(N-j)a^2}\sin(2\alpha_r)\alpha\leq0
\end{equation}
for $0\leq r\leq r_{opt}$ due to $a(r)=\sin\alpha_r\geq a(0)=\sqrt{j/N}$,
which means that $\Cr(\rho)$ is a decreasing function of $r$.
On the contrary, the success probability $P(r)$ is a increasing function for $0\leq r\leq r_{opt}$.
Moreover, the coherence achieves the minimal value while the probability of success reaches the maximal value 1.
That is to say, the improvement of success probability depends on the QCD, see Fig. 2.

\begin{figure}[ht]
 \centering
 \includegraphics[height=7cm]{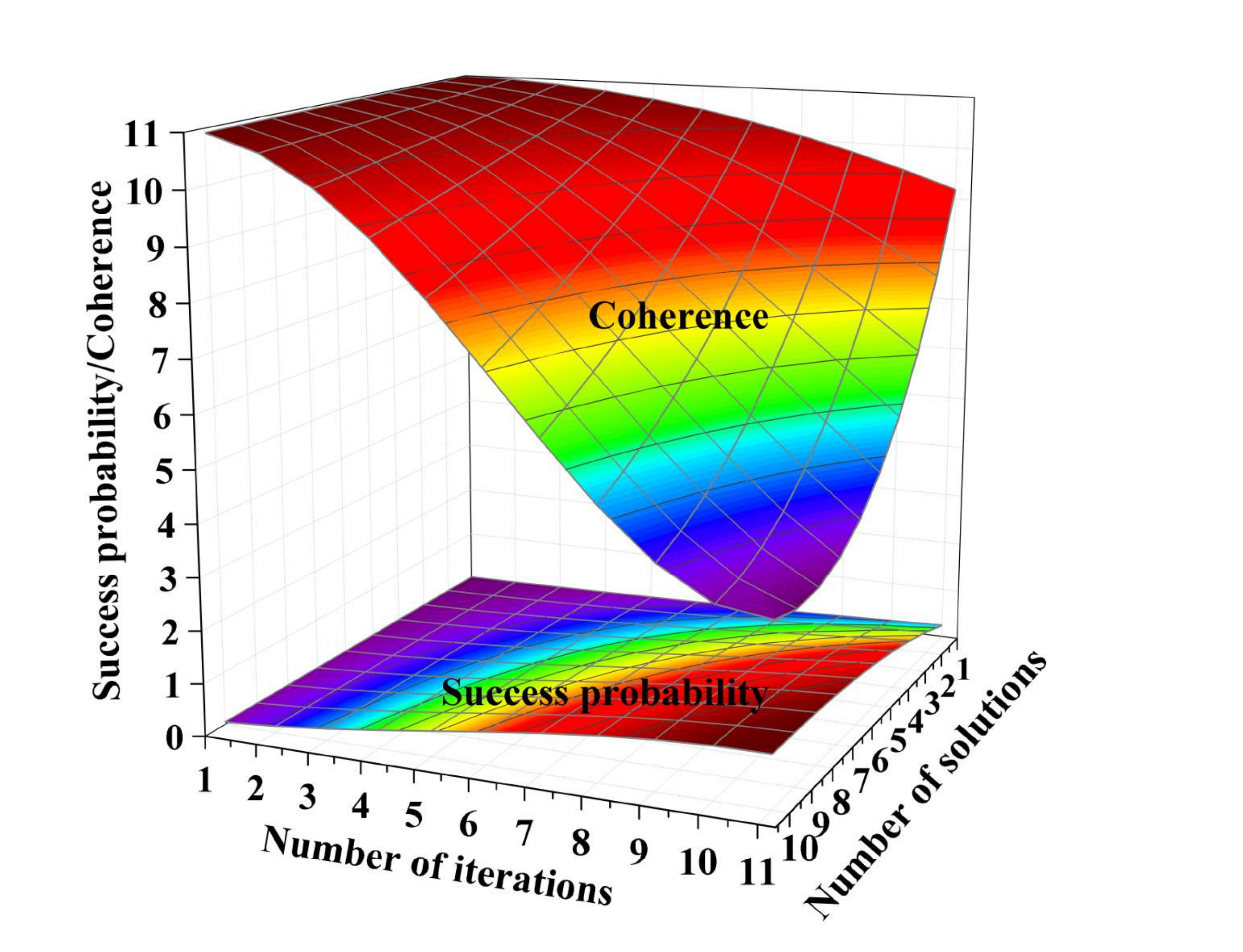}
 \caption{(Color online) Evolution of coherence in GA for the whole 11-qubit system with $j$ (from 1 to 10) solutions.}
 \label{fig:algorithm}
\end{figure}

It is possible to express the coherence $\Cr(\rho)$ as a function of the success probability $P$.
Due to the fact that $P=a^2$, the coherence becomes
\begin{equation}\label{C-r-P}
\Cr(\rho)=H(P)+\log_2(N-j)+P\log_2\frac{j}{N-j}.
\end{equation}
Actually, the GA is usually applied in the situation that a few solutions in a huge database.
Under this condition ($j\ll N$ and $N\gg1$), $H(P)$ can be omitted compared with $\log_2(N-j)$ and then Eq. (\ref{C-r-P}) takes the following form
\begin{equation}
\Cr(\rho)\simeq -P\log_2\frac{N}{j}+\log_2N,
\end{equation}
which is a linear function of $P$.
The ability of coherence in enhancing the success probability can be quantified as cost performance $w$,
\begin{equation}
w=-\frac{\ud P}{\ud\Cr}=\frac{1}{\log_2\frac{N}{j}}.
\end{equation}
Clearly, the cost performance is related to a constant $j/N$, which represent the ratio of number of solutions to the scale of database.

\subsection{The $l_1$ norm of coherence}
The $l_1$ norm of coherence is a very intuitive quantification
which comes from a simple fact that coherence is linked with the off-diagonal elements of considered quantum states.
The expression of the $l_1$ norm of coherence is defined as \cite{Baumgratz}
\begin{equation}
\Cl(\rho)=\sum_{i\neq j}|\rho_{ij}|.
\end{equation}
By employing this equation, we have the coherence dynamics in GA
\begin{equation}
\Cl(\rho)=(\sqrt{j}\sin\alpha_r+\sqrt{N-j}\cos\alpha_r)^2-1,
\end{equation}
when $0\leq r\leq r_{opt}$. Using Eq. (\ref{P}), the $l_1$ norm of coherence can be rewritten as a function of $P$
\begin{equation}
\Cl(\rho)=(\sqrt{jP}+\sqrt{(N-j)(1-P)})^2-1.
\end{equation}
In the asymptotic limits $j\ll N$ and $N\gg1$, the $l_1$ norm of coherence takes the simple form
\begin{equation}
\Cl(\rho)\simeq-NP+N.
\end{equation}
The same phenomenon that the success probability depends on the QCD is also existed under the $l_1$ norm of coherence measure.
From this perspective, we say that the QCD is of great significance in GA.
And the cost performance $w$ equals to $1/N$.

\section{COHERENCE DEPLETION IN GENERALIZED GROVER SEARCH ALGORITHM}\label{section2}
In Ref. \cite{GGA}, Grover search algorithm was generalized to deal with arbitrary initial complex amplitude distributions.
The only difference between GA and generalized grover search algorithm (GGA) is there are no initialization step in GGA.
Thus the GGA includes the following steps\\
(1) Use any initial amplitude distribution of a system which does not need to be initialized to the uniform distribution.\\
(2) Repeat the following two steps $r$ times: (i) Rotate the solutions by a phase of $\pi$ radians.
(ii) Rotate all states according to the average amplitude of all states by $\pi$.\\
(3) Measure the resulting state in the optimal time $r_{opt}$.\\
We denote the amplitudes of solutions by $k_i(r)$, $i=1,...,j$, and non-solutions by $l_i(r)$, $i=j+1,...,N $.
Let the average amplitudes over solutions and over non-solutions are represented respectively by
\begin{eqnarray}
&&\bar{k}(r)=\frac{1}{j}\sum_{i=1}^{j}k_i,\\
&&\bar{l}(r)=\frac{1}{N-j}\sum_{i=j+1}^Nl_i.
\end{eqnarray}
The success probability in optimal measurement time was founded by Biham \emph{et al.} \cite{GGA},
\begin{equation}\label{P-GGA}
P_{max}^{GGA}=1-(N-j)\sigma_l^2.
\end{equation}
where $\sigma_l^2=1/(N-j)\sum_{i=j+1}^N|l_i(r_{opt}^{GGA})-\bar{l}(r_{opt}^{GGA})|^2$, and the optimal measurement time is given by
\begin{equation}
r_{opt}^{GGA}=(\frac{\pi}{2}-\beta)/\omega
\end{equation}
with $\cos \omega=(1-2j/N)$ and $\tan \beta=\sqrt{j/(N-j)}$ $\bar{k}(0)/\bar{l}(0)$.
The dynamics of the amplitudes are described by \cite{GGA}
\begin{eqnarray}\label{GGA-rho1}
&&k_i(r)=\bar{k}(r)+k_i(0)-\bar{k}(0),\\\label{GGA-rho2}
&&l_i(r)=\bar{l}(r)+(-1)^r[l_i(0)-\bar{l}(0)].
\end{eqnarray}

Now let us consider the such initial state
\begin{equation}\label{GGA-rho-initial}
\ket{\phi_0}=\phi_0\ket{0}+\phi_1\ket{1}+\frac{1}{\sqrt{N}}\sum_{x=2}^{N-1}\ket{x},
\end{equation}
where $\phi_0^2+\phi_1^2=2/N$, $\phi_0$, $\phi_1 \in \mathbb{R}$ and $\ket{0}$, $\ket{1}$ are solutions.
Without loss of generality,
we assume that $\phi_0\leq \phi_1$.
From Eqs. (\ref{P-GGA}) and (\ref{GGA-rho-initial}), it follows immediately that the success probability of these kind of states can reach the maximum value 1,
and corresponding states can be written as
\begin{equation}\label{GGA-rho-opt}
\ket{\phi_{opt}}=k_1\ket{0}+k_2\ket{1},
\end{equation}
with $k_1=\sqrt{(N-2)/(2N)+1/4(\phi_0+\phi_1)^2}+1/2(\phi_0-\phi_1)$ and $k_2=\sqrt{(N-2)/(2N)+1/4(\phi_0+\phi_1)^2}-1/2(\phi_0-\phi_1)$.
By using Eqs. (\ref{C-r}), (\ref{GGA-rho-initial}) and (\ref{GGA-rho-opt}), the QCD of these kind of states in GGA is
\begin{eqnarray}
\Delta\Cr^{GGA}&\equiv& \Cr(\ket{\phi_0}\bra{\phi_0})-\Cr(\ket{\phi_{opt}}\bra{\phi_{opt}})\n
&=&-\phi_0^2\log_2\phi_0^2-\phi_1^2\log_2\phi_1^2\n
& &+\frac{N-2}{N}\log_2N-H(k_1^2),
\end{eqnarray}
where $H$ is the binary Shannon entropy function. Both $\Delta\Cr^{GGA}$ and $r_{opt}^{GGA}$ are increased with the decrease of $\phi_0$, see Fig. 3.
\begin{figure}
 \centering
 \includegraphics[height=5.5cm]{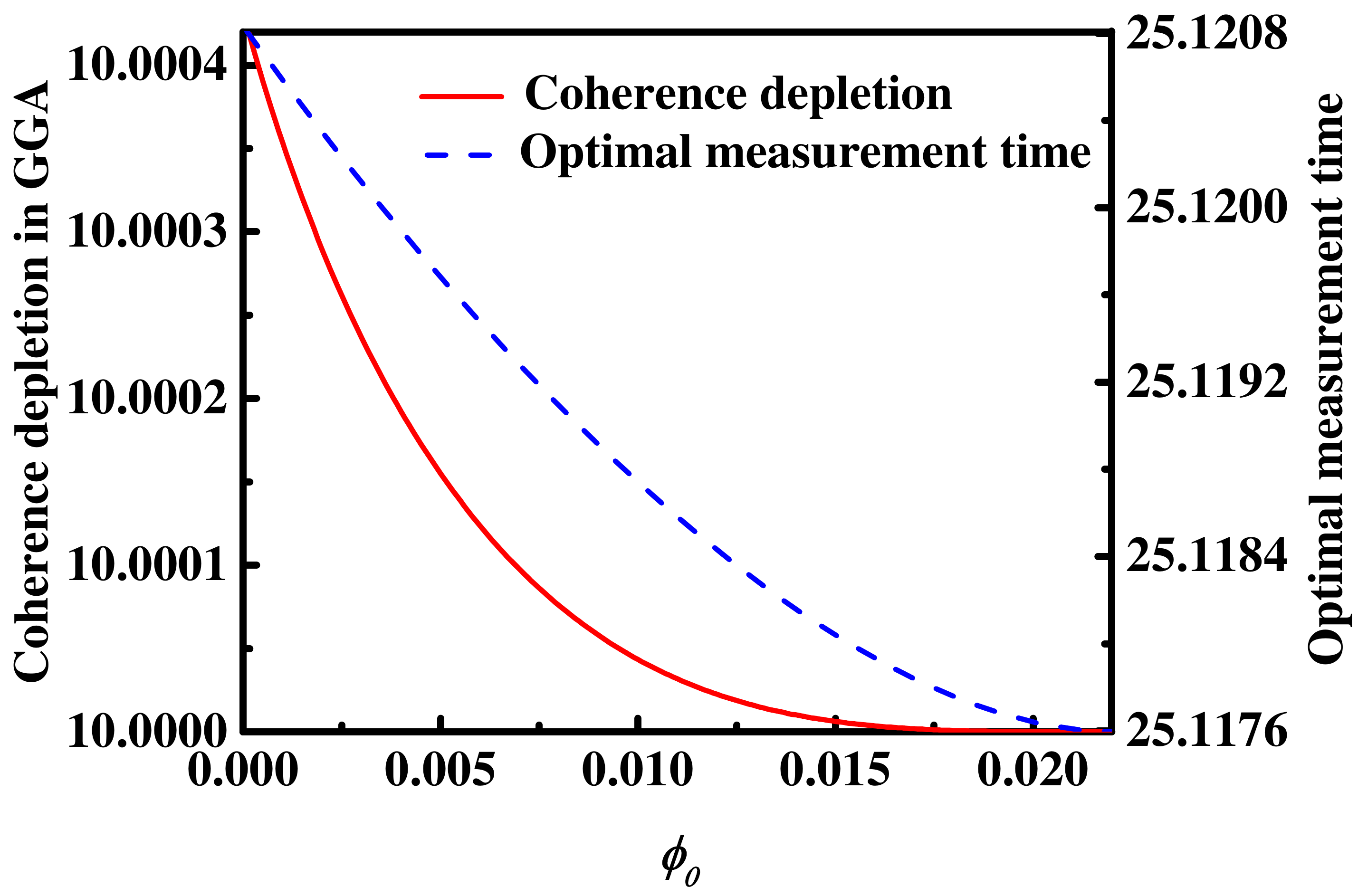}
 \caption{(Color online) Coherence depletion $\Delta\Cr^{GGA}$ versus optimal measurement time $r_{opt}^{GGA}$ for the initial states of $\ket{\phi_0}$ in GGA.}
 \label{fig:algorithm}
\end{figure}
It means that the optimal measurement time depends on the QCD for this kind of states in GGA.
In other words, the smaller the optimal measurement time is, the smaller the QCD is.

\section{OTHER QUANTUM CORRELATIONS IN GROVER SEARCH ALGORITHM}\label{section3}
In this section, we only consider the simplest situation of single solution ($j=1$) for convenience,
 which benefits to capturing the essence of other quantum resources dynamics in GA.
Without loss of generality,
we assume that the solution is located at $\ket{0}$ and the density matrix of states generated by GA (Eq. (\ref{rho-r})) has the following form
\begin{equation}\label{rho-1}
\rho(r)=\left(
  \begin{array}{ccccc}
    a^2 & ab  & ab  & ab  & \cdots \\
    ab  & b^2 & b^2 & b^2 & \cdots \\
    ab  & b^2 & b^2 & b^2 & \cdots \\
    ab  & b^2 & b^2 & b^2 & \cdots \\
    \vdots & \vdots & \vdots & \vdots & \ddots \\
  \end{array}
\right)
_{N\times N}.
\end{equation}

\subsection{Entanglement in Grover search}
Entanglement is widely considered as the main undertaker for quantum computational speed-up though the role of entanglement is not clear. Here we use concurrence, a widely-accepted entanglement measure, to investigate the behavior of entanglement during the GA.

The reduced matrix of any two qubits takes the following form
\begin{equation}\label{Reduce-rho-2}
\rho_2=
\left(
  \begin{array}{cccc}
    \Omega_0 & \Omega_1 & \Omega_1 & \Omega_1 \\
    \Omega_1 & \Omega_2 & \Omega_2 & \Omega_2 \\
    \Omega_1 & \Omega_2 & \Omega_2 & \Omega_2 \\
    \Omega_1 & \Omega_2 & \Omega_2 & \Omega_2 \\
  \end{array}
\right),
\end{equation}
where $\Omega_0=a^2+(\frac{N}{4}-1)b^2$, $\Omega_1=ab+(\frac{N}{4}-1)b^2$ and $\Omega_2=\frac{N}{4}b^2$.
The concurrence of arbitrary two-qubit states is defined in Ref \cite{Wootters} and is calculated as follows
\begin{equation}\label{Concurrence-2}
E_2(\rho)=\max\{0,\lambda_1-\lambda_2-\lambda_3-\lambda_4\},
\end{equation}
where $\lambda_is$ are square roots of the eigenvalues of matrix $\rho\tilde{\rho}$ in decreasing order, $\lambda_1\geq\lambda_2\geq\lambda_3\geq\lambda_4$. Here $\tilde{\rho}=(\sigma_y\otimes\sigma_y)\rho^*(\sigma_y\otimes\sigma_y)$,
where $\sigma_y$ is Pauli matrix $\begin{pmatrix} 0 & -i \\ i &  0 \end{pmatrix}$, and $\rho^*$ is the conjugation of $\rho$.
According to Eqs. (\ref{Reduce-rho-2}) and (\ref{Concurrence-2}),
the expression of concurrence between any two qubits $\rho_2$ in GA can be obtained \cite{Fang}
\begin{equation}
E_2(\rho_2)= 2|\Omega_1-\Omega_2|=2|ab-b^2|.
\end{equation}
The  behavior of pairwise entanglement in the case of $n=11$ is displayed in Fig. 4.
The pairwise entanglement firstly increases to the maximal value and then decreases to almost zero when the optimal number of iterations is reached.
\begin{figure}
 \centering
 \includegraphics[height=6cm]{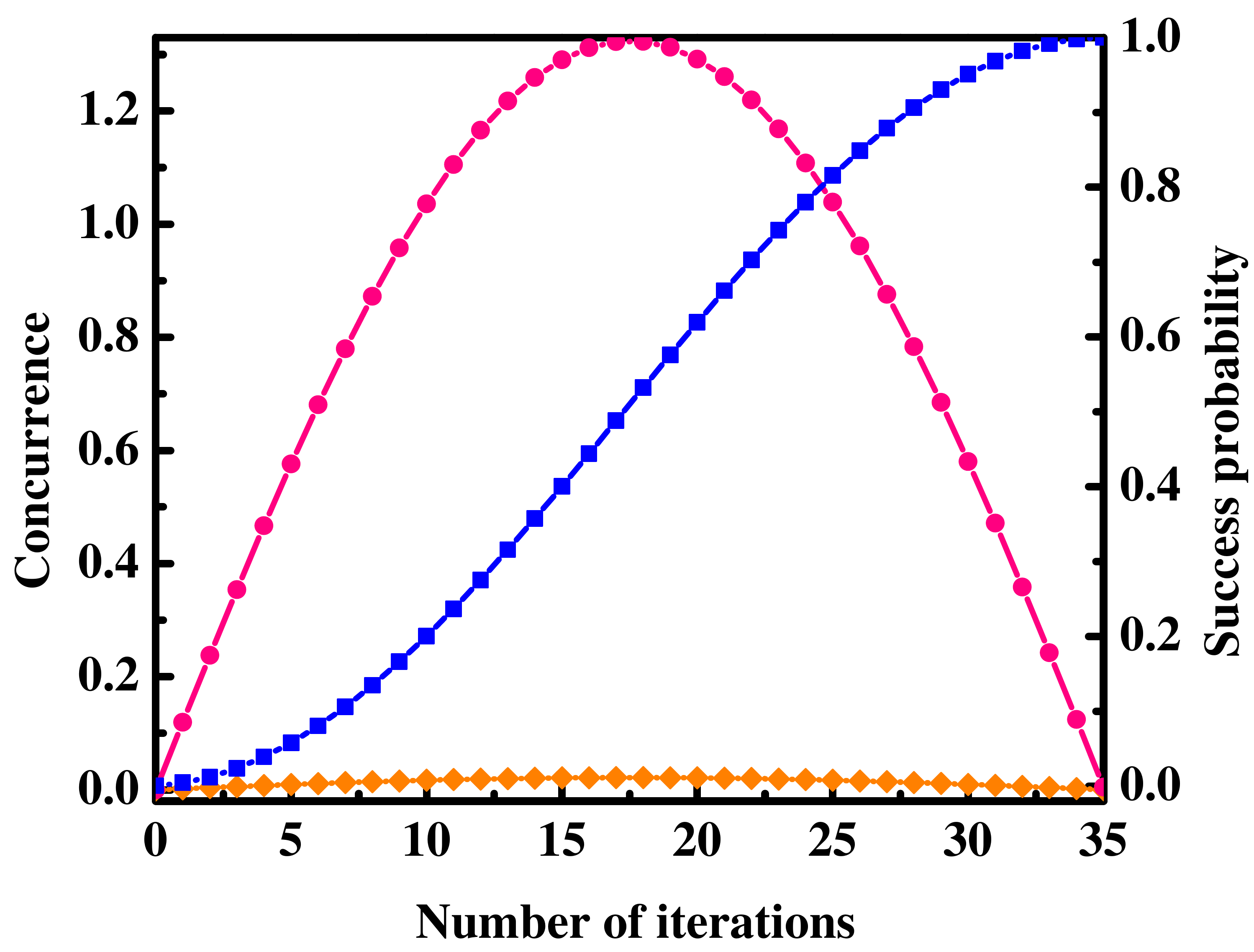}
 \caption{(Color online) The evolutions for the entanglement in the case of 11-qubit system. The pairwise entanglement is depicted by orange diamonds while the entanglement of the whole 11-qubit system by pink points. The blue squares represent the success probability.}
 \label{fig:algorithm}
\end{figure}

Now let's consider the  multipartite entanglement of  $n$-qubit system, which may better depict the behavior of $P(r)$.
The concurrence of $n$-qubit states is introduced in Ref \cite{Carvalho}
\begin{equation}
E_n(\psi)=\frac{2}{\sqrt{N}}\sqrt{(N-2)\bra{\psi}\psi\rangle^2-\sum_{\beta}\Tr\rho_{\beta}^2},
\end{equation}
where $N=2^n$ and $\beta$ labels $(N-2)$ different reduced density matrices;
i.e., there are $C^k_N$ different terms when tracing over $k$ different subsystems from the $n$-qubit system.
Note that the concurrence for $n$-qubit states used is upper bound.
From Eq. (\ref{rho-1}), we have reduced matrix for any $k$-qubit
\begin{widetext}
\begin{equation}\label{rho-reduced}
\rho_k=\left(
  \begin{array}{ccccc}
    a^2+(2^{n-k}-1)b^2 & ab+(2^{n-k}-1)b^2  & ab+(2^{n-k}-1)b^2  & ab+(2^{n-k}-1)b^2  & \cdots \\
    ab+(2^{n-k}-1)b^2  & 2^{n-k}b^2         & 2^{n-k}b^2         & 2^{n-k}b^2         & \cdots \\
    ab+(2^{n-k}-1)b^2  & 2^{n-k}b^2         & 2^{n-k}b^2         & 2^{n-k}b^2         & \cdots \\
    ab+(2^{n-k}-1)b^2  & 2^{n-k}b^2         & 2^{n-k}b^2         & 2^{n-k}b^2         & \cdots \\
    \vdots & \vdots & \vdots & \vdots & \ddots \\
  \end{array}
\right)
_{2^k\times 2^k}.
\end{equation}
\end{widetext}
Thereby concurrence of the whole $n$-qubit system can be expressed as
\begin{equation}
E_n=\frac{2}{\sqrt{N}}\sqrt{(N-2)-\sum_{k=1}^{n-1}C_n^k\Tr\rho_k^2}.
\end{equation}
Substitute Eq. (\ref{rho-reduced}) into the above equation, we have
\begin{eqnarray}
E_n&=&\frac{2}{\sqrt{2^n}}[2^n-2-(4\times3^n-2^{n+3}+4)a^2b^2\n
& &-(8^n+4\times3^n-3\times2^{2n+1}+3\times2^n-2)b^4\n
& &-(2^n-2)a^4-4(4^n-2\times3^n+2^n)ab^3]^{\frac{1}{2}}.
\end{eqnarray}
By virtue of this equation, we present the behavior of multipartite entanglement of $n$-qubit system in the case that $n=11$,
which is similar with the pairwise entanglement (see Fig. 4).

\subsection{Discord in Grover search}
Discord was introduced in Ref. \cite{Ollivier} to quantify quantum correlation,
which is viewed as the difference between total correlation and classical correlation
\begin{equation}\label{Discord-def}
\D(\rho)=\I(\rho)-\C(\rho),
\end{equation}
where $\I$ and $\C$ represent the total correlation and classical correlation, respectively.
In Ref. \cite{Groisman}, the total correlation between two systems $A$ and $B$ is defined by the minimal amount of noise
, which is wanted to destroy all the correlation between them.
The total correlation is equal to the quantum mutual information
\begin{equation}\label{Total-Correlation}
\I(\rho_{AB})=S(\rho_A)+S(\rho_B)-S(\rho_{AB}),
\end{equation}
where $\rho_{A(B)}=\Tr_{B(A)}\rho_{AB}$.
The classical correlation was proposed in Ref. \cite{Henderson} as the maximum information we can obtain from $A$ by measuring $B$.
Under projective measurements $\{\prod_i\}$, the classical correlation can be written as
\begin{equation}\label{Concurrence-def}
\C(\rho)=\max_{\{\prod_i\}}\{S(\rho_A)-\sum_ip_iS(\rho_{A|i})\},
\end{equation}
where $p_i=\Tr_{AB}(I\otimes\prod_i)\rho_{AB}(I\otimes\prod_i)$ and $\rho_{A|i}=1/p_i\Tr_B(I\otimes\prod_i)\rho_{AB}(I\otimes\prod_i)$.
Put Eqs. (\ref{Total-Correlation} and (\ref{Concurrence-def}) into Eq. (\ref{Discord-def}), then
\begin{equation}
\D(\rho)=\min_{\{\prod_i\}}\sum_i[p_iS(\rho_{A|i})+S(\rho_B)-S(\rho_{AB})].
\end{equation}
We choose the bipartite discord to analyse discord dynamics in GA.
The projective measurement can be parameterized via $0\leq\theta\leq\pi$ and $0\leq\phi\leq2\pi$ in the form of $\{\cos\theta\ket{0}+e^{i\phi}\sin\theta\ket{1}$, $e^{-i\phi}\sin\theta\ket{0}-\cos\theta\ket{1}\}$.
Using the exact diagonalization method, we calculate pairwise discord in the case of 11-qubit system.
The Fig. 5 shows that the behavior of pairwise discord is similar to the entanglement.
\begin{figure}[ht]
 \centering
 \includegraphics[height=6cm]{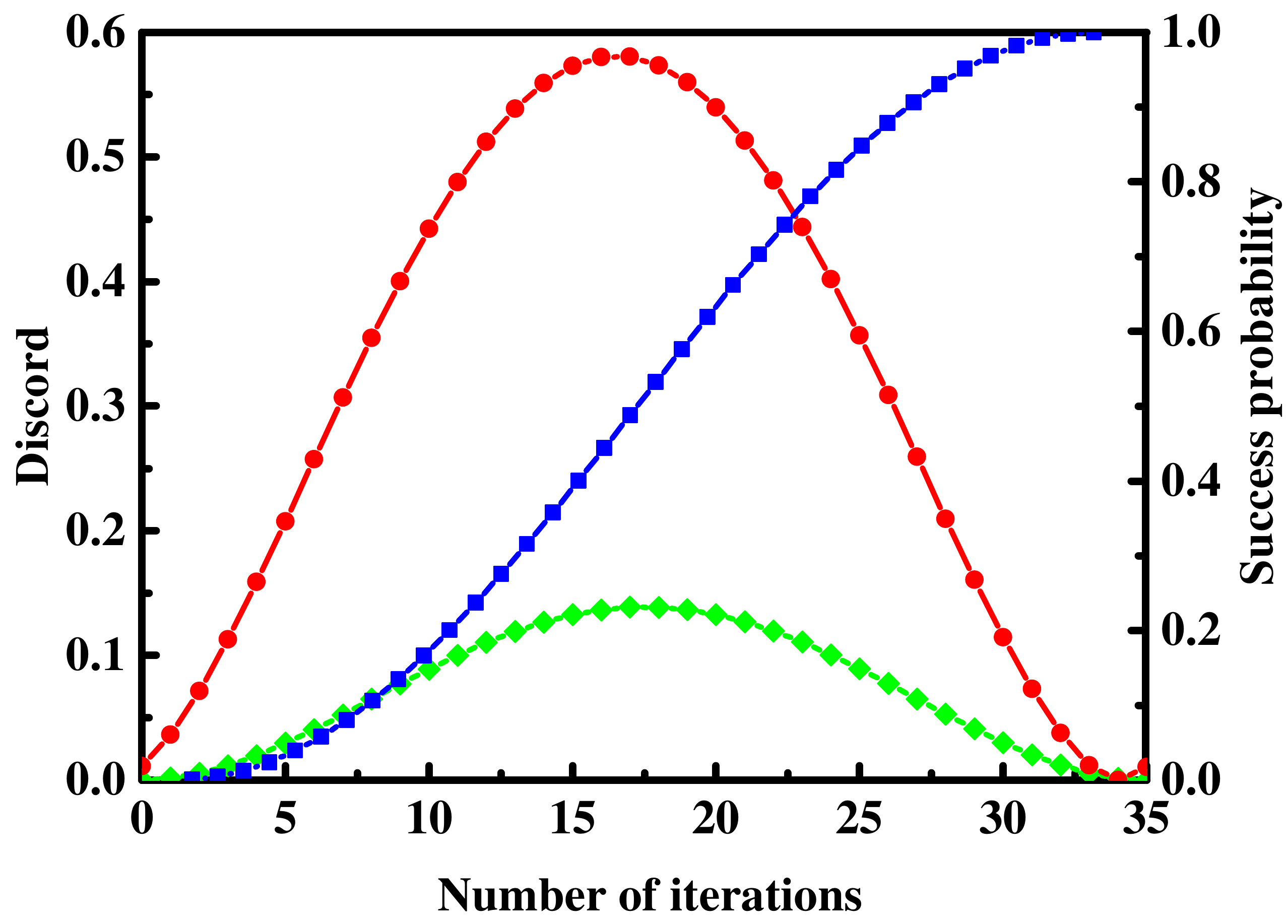}
 \caption{(Color online)  The evolutions for the discord in the case of 11-qubit system. The pairwise discord is depicted by green diamonds while the genuine quantum correlation of the whole 11-qubit system by red points. The blue squares represent the success probability.
 }
 \label{fig:algorithm}
\end{figure}

In Ref. \cite{Genuine Discord}, a quantifier for genuine multipartite quantum correlation was proposed based on relative entropy.
For tripartite pure states $\rho_{ABC}$, the genuine quantum correlation $\D^{(3)}$ is equal to half of genuine total correlation $T^{(3)}$, namely
\begin{equation}\label{D3}
\D^{(3)}(\rho_{ABC})=\frac{T^{(3)}(\rho_{ABC})}{2}.
\end{equation}
Here $T^{(3)}$ is defined as the difference between total correlation $T$ and the maximum among the bipartite correlation $T^{(2)}$
\begin{equation}
T^{(3)}(\rho_{ABC})=T(\rho_{ABC})-T^{(2)}(\rho_{ABC}),
\end{equation}
where $T(\rho_{ABC})=S(\rho_A)+S(\rho_B)+S(\rho_C)-S(\rho_{ABC})$ and $T^{(2)}(\rho_{ABC})=\max\{\I(\rho_{AB}),\I(\rho_{AC}),\I(\rho_{BC})\}$.
Defined in this way, $T^{(3)}$ is the shortest distance to a state without tripartite correlations based on relative entropy.
For pure states of $n$ qubits, genuine $n$-partite quantum correlation can also be expressed as \cite{Genuine Discord}
\begin{equation}
\D^{(n)}(\rho)=\frac{T^{(n)}(\rho)}{2},
\end{equation}
where $T^{(n)}(\rho)=S(\rho||\sigma)$ and $\sigma$ is a product state making $S(\rho||\sigma)$ minimum.
Besides, Modi \emph{et al.} founded that $\sigma$ is the reduced states of $\rho$ in the product form.
According to Eq. (\ref{rho-reduced}), we obtain
\begin{eqnarray}
T^{(n)}(\rho)&=&\min_{\sum_ik_i=n}S(\rho||\bigotimes_i\rho_{k_i})\n
&=&\min_{\sum_ik_i=n}[-S(\rho)-\Tr(\rho\log_2\bigotimes_i\rho_{k_i})]\n
&=&\min_{\sum_ik_i=n}[-\sum_i\Tr(\rho_{k_i}\log_2\rho_{k_i})]\n
&=&\min_{\sum_ik_i=n}\sum_iS(\rho_{k_i})
\end{eqnarray}
since $\rho$ in GA is a pure state.
By using Lagrangian multiplier method, above equation is simplify into
\begin{eqnarray}
T^{(n)}(\rho)
&=&S(\rho_1)+S(\rho_{n-1})\n
&=&2S(\rho_1),
\end{eqnarray}
where $\rho_1$ is a reduced state of any single qubit in GA
\begin{equation}\label{rho-reduced-1}
\rho_1=\left(
  \begin{array}{cc}
    a^2+(2^{n-1}-1)b^2 & ab+(2^{n-1}-1)b^2  \\
    ab+(2^{n-1}-1)b^2  & 2^{n-1}b^2        \\
  \end{array}
\right).
\end{equation}
Thus, the dynamics of genuine quantum correlation in GA becomes
\begin{equation}
D^{(n)}=S(\rho_1)=H(\frac{1+\sqrt{\Delta}}{2}),
\end{equation}
where $\Delta=1-4(2^{n-1}-1)(ab-b^2)^2$.
The Fig. 5 depicts the behavior of genuine quantum correlation of the whole 11-qubit system in GA.

\subsection{Nonlocality in Grover search}
Nonlocality is another manifestation of nonclassical correlation
which tells us that reproducing the predictions of quantum theory by considering local hidden variables (LHV) is impossible.
It is well known that the entanglement is necessary for the existence of nonlocality but nonlocality is not necessary for entanglement \cite{Wiseman}.
We are interested about whether nonlocality appears in GA or not.
Unfortunately, there is a lack of necessary and sufficient criterions or suitable measurements for nonlocality.
Violating the CHSH (Clauser, Horne, Shimony and Holt) inequality provides a powerful tool to recognize the nonlocality of two-qubit systems.
Consequently, we choose the CHSH inequality to investigate the nonlocality of any two qubits during the Grover search.

The CHSH inequality is described as \cite{CHSH}
\begin{equation}
|\langle \B_{CHSH}\rangle|=|\Tr(\B_{CHSH}\rho)|\leq2,
\end{equation}
where
\begin{equation}
\B_{CHSH}=\vec{a}\cdot\vec{\sigma_1}\otimes(\vec{b}+\vec{b'})\cdot\vec{\sigma_2}+\vec{a'}\cdot\vec{\sigma_1}\otimes(\vec{b}-\vec{b'})\cdot\vec{\sigma_2}
\end{equation}
and $\vec{a}, \vec{a'}, \vec{b}, \vec{b'}$ are unit vectors in $\mathbb{R}^3$.
In Ref. \cite{Horodecki}, a theorem, that a two-qubit system violates the CHSH inequality if and only if $M(\rho)>1$, has been given.
Note that obeying the CHSH inequality does not mean that the system is local.
Here, $M(\rho)=\max_{i\neq j}\{u_i +u_j\}$ with $u_i$ being the three eigenvalues of the matrix $T^TT$,
where $T = T_{ij} = \Tr\rho(\sigma_i\otimes\sigma_j)$ is the correlation matrix.
The correlation matrix for $\rho_2$ is given by
\begin{equation}
T=\begin{pmatrix} 2\Omega_1+2\Omega_2 & 0 & 2\Omega_1-2\Omega_2 \\ 0 & 2\Omega_2-2\Omega_1 & 0 \\ 2\Omega_1-2\Omega_2 & 0 & \Omega_0-\Omega_2 \\ \end{pmatrix}
\end{equation}
and the corresponding eigenvalues are $\lambda_1=2\Omega_2-2\Omega_1$, $\lambda_2=(\Omega_0+2\Omega_1+\Omega_2-\sqrt{\bigtriangleup})/2$ and $\lambda_3=(\Omega_0+2\Omega_1+\Omega_2+\sqrt{\bigtriangleup})/2$ with $\bigtriangleup=\Omega_0^2+20\Omega_1^2+25\Omega_2^2-4\Omega_0\Omega_1-6\Omega_0\Omega_2-20\Omega_1\Omega_2$.
Therefore, we have
\begin{eqnarray}
M(\rho_2)=\left\{
            \begin{array}{ll}
              \lambda_2^2+\lambda_3^2, & \hbox{$\lambda_1\leq\lambda_2$;} \\
              \lambda_1^2+\lambda_3^2, & \hbox{$\lambda_1>\lambda_2$.}
            \end{array}
          \right.
\end{eqnarray}
In the asymptotic limits $N\gg1$, we have $\lambda_1\leq\lambda_2$ and
\begin{eqnarray}
M(\rho_2)&=&\lim_{N\rightarrow\infty}\lambda_2^2+\lambda_3^2=\lim_{N\rightarrow\infty}\frac{(\Omega_0+2\Omega_1+\Omega_2)^2+\triangle}{2}\n
&=&\lim_{N\rightarrow\infty}\Omega_0^2+12\Omega_1^2+13\Omega_2^2-2\Omega_0\Omega_2-8\Omega_1\Omega_2\n
&=&1-2\sin^2(\frac{2r+1}{2}\alpha)\cos^2(\frac{2r+1}{2}\alpha)\n
&\leq&1,
\end{eqnarray}
which means that the pairwise nonlocality does not exist in this limit case.

Next we will discuss genuine tripartite nonlocality of reduced tripartite states in the GA by using the Svetlichny's inequality.
The violation of Svetlichny's inequality means that the correlations cannot be simulated by a hybrid nonlocal-local ensemble \cite{Svetlichny},
thus the correlation is genuine tripartite nonlocality.
Svetlichny's inequality is in the form of \cite{Svetlichny}
\begin{equation}
|<\B_S>|=|\Tr(\B_S\rho)|\leq4.
\end{equation}
where $\B_S$ is the Svetlichny's operator and defined as
\begin{eqnarray}\label{B-S-inequality}
\B_S&=&A[(B+B')C+(B-B')C']\n
&&+A'[(B-B')C-(B+B')C'].
\end{eqnarray}
Here the measurements are spin projections onto unit vectors: $A=\vec{a}\cdot\vec{\sigma_1}$ ($A'=\vec{a'}\cdot\vec{\sigma_1}$) on the first qubit,
$B=\vec{b}\cdot\vec{\sigma_2}$ ($B'=\vec{b'}\cdot\vec{\sigma_2}$) on the second qubit,
and $C=\vec{c}\cdot\vec{\sigma_3}$ ($C'=\vec{c'}\cdot\vec{\sigma_3}$) on the third qubit.
By defining $\vec{b}+\vec{b'}=2\vec{d}\cos t$ and $\vec{b}-\vec{b'}=2\vec{d'}\sin t$ ($\vec{d}\cdot\vec{d'}=0$),
$\B_S$ can be further simplified as
\begin{eqnarray}\label{B-S}
|\langle\B_S\rangle|&=&2|(\langle AD'C'\rangle\sin t-\langle A'DC'\rangle\cos t)\n
&&+(\langle A'D'C\rangle\sin t+\langle ADC\rangle\cos t)|\n
&\leq&2(\sqrt{\langle AD'C'\rangle^2+\langle A'DC'\rangle^2}\n
&&+\sqrt{\langle A'D'C\rangle^2+\langle ADC\rangle^2})\n
\end{eqnarray}
where $D=\vec{d}\cdot\vec{\sigma_2}$ and $D'=\vec{d'}\cdot\vec{\sigma_2}$.

Any tripartite states can be expressed as
\begin{eqnarray}
\rho_3&=&\frac{1}{8}(I+\sum_i e_i\sigma_i\otimes I\otimes I+\sum_if_iI\otimes\sigma_i\otimes I\n
&&+\sum_ig_iI\otimes I\otimes \sigma_i+\sum_{ij}M^a_{ij}I\otimes\sigma_i\otimes\sigma_j\n
&&+\sum_{ij}M^b_{ij}\sigma_i\otimes I\otimes\sigma_j+\sum_{ij}M^c_{ij}\sigma_i\otimes\sigma_j\otimes I\n
&&+\sum_{ijk}T_{ijk}\sigma_i\otimes\sigma_j\otimes\sigma_k),
\end{eqnarray}
and
\begin{equation}\label{T}
T_{ijk}=\Tr(\sigma_i\otimes\sigma_j\otimes\sigma_k\rho_3).
\end{equation}
In the asymptotic limits $N\gg1$, by using Eqs. (\ref{rho-reduced}) and (\ref{T}),
the $T$ of reduced tripartite states generated by GA has only two nonzero elements: $T_{111}=(\cos^2\alpha_r)/4$ and $T_{333}=(\sin^2\alpha_r)/8$.
Let $\tilde{T}_{ij}=\sum_{k}T_{ijk}c'_k$, it gives
\begin{eqnarray}
\langle AD'C'\rangle^2+\langle A'DC'\rangle^2&=&\max\{[\vec{a}\cdot(\tilde{T}\vec{d'})]^2+[\vec{a'}\cdot(\tilde{T}\vec{d})]^2\}\n
&=&\max\{||\tilde{T}\vec{d'}||^2+||\tilde{T}\vec{d}||^2\}\n
&=&v_1+v_2,
\end{eqnarray}
where $v_1$ and $v_2$ are two greater eigenvalues of $\tilde{T}^T\tilde{T}$, $v_1=c_1'^2/16\cos^4\alpha_r$ and $v_2=c_3'^2/64\sin^4\alpha_r$.
Thus,
\begin{equation}\label{Max-1}
\langle AD'C'\rangle^2+\langle A'DC'\rangle^2\leq1.
\end{equation}
Similarly, we can also obtain
\begin{equation}\label{Max-2}
\langle A'D'C\rangle^2+\langle ADC\rangle^2\leq1.
\end{equation}
According to Eqs. (\ref{B-S}), (\ref{Max-1}) and (\ref{Max-2}), we have
\begin{equation}
|\Tr(\B_S\rho_3)|\leq4,
\end{equation}
which means genuine tripartite nonlocality is not detected by using Svetlichny's inequality.

\section{Conclusions}\label{section5}
In this work, we have systematically studied the evolutions of coherence and other typical quantum correlations in the process of Grover search.
Eventually, we find that both success probability in GA and optimal measurement time in GGA can be directly related to a scalar function of state, QCD.
By using the relative entropy measure of coherence, we show that
the improvement of success probability relies on the coherence depletion for any number of solutions in GA.
Explicitly, in the limit case of a few searcher items $j\ll N$ and large database $N\gg1$, the cost performance about coherence in enhancement the success probability is associated with the ratio of number of searched solutions to the scale of database, $j/N$.
The same phenomenon also exists by using the $l_1$ norm of coherence and corresponding cost performance equals to $1/N$.
In GGA, we discover a class of states where the required optimal measurement time increases with the QCD.

For pairwise entanglement, multipartite entanglement, pairwise discord and genuine quantum correlation,
they are always present during the whole process of GA.
The behaviors of them generally start from zero then reach the maximum and decrease to almost zero.
But we fail to connect them with success probability.
Moreover, in the limit case, the nonlocalities with respect to two-qubit and three-qubit systems
have not been detected during GA by using CHSH type Bell inequality and Svetlichny's inequality.

Our research exhibits the significance of QCD in Grover search algorithm,
contributing to the resource theory of quantum coherence and
providing deep insights into the role of coherence in quantum algorithms.
On one hand, QCD increases the success probability in GA.
On the other hand, a smaller amount of QCD decreases the required optimal measure time in GGA.
Therefore, the coherence can be viewed as a potential resource in Grover search algorithm.
Our method is also worth applying to investigate QCD in other quantum information process,
such as Shor's algorithm, teleportation and so on.

\section{Acknowledgements}
We thank L. C. Zhao, J. X. Hou and Y. H. Shi for their valuable discussions.
This work was supported by the NSFC (Grant No.11375141, No.11425522, No.91536108 and No.11647057),
the special research funds of shaanxi province department of education (No.203010005),
Northwest University scientific research funds (No.338020004)
and the double first-class university construction project of Northwest
University.

\end{document}